\documentclass[sn-mathphys]{sn-jnl}

\usepackage[english]{babel}
 \usepackage{amsmath}
\usepackage{amssymb}
\usepackage{graphicx}
\usepackage{bm}
\usepackage{slashed}
\usepackage{yfonts}
\usepackage{marvosym}
\usepackage{wasysym}

\newcommand{\myskip}[1]{}

\newcommand{\Schw}{{\rm S}}

\newcommand{\pg }{{\rm pg}}

\newcommand{\bSe}{\bar \Se  }

\newcommand{\sho}{{\it sho}}

\newcommand{\Mc}{M_c}
\newcommand{\qi}{q_i}

\newcommand{\cQc}{{\cal Q}_c}
\newcommand{\rhom}{\rho_\lambda}

\newcommand{\atu}{\tau}

\newcommand{\pplus}{\hspace{-0.5mm}+\hspace{-0.5mm}}
\newcommand{\mmin}{\hspace{-0.5mm}-\hspace{-0.5mm}}

\newcommand{\tot}{{\rm tot}}

\newcommand{\half}{{\frac{1}{2}}}

\renewcommand{\p}{\partial}   
\newcommand{\onedot}{\,\,\,}   
\newcommand{\twodots}{\,\,\,\,}   
\newcommand{\threedots}{\,\,\,\,\,\,}   
\newcommand{\ed}{\onedot}
\newcommand{\ednu}{{\onedot\nu}}

\newcommand{\td}{\twodots}   
\newcommand{\dd}{\threedots}

\newcommand{\Om}{\Omega}

\renewcommand{\th}{{\theta}}

\newcommand{\vth}{{\vartheta}}

\newcommand{\om}{\omega}

\newcommand{\gam}{\gamma}   
   
\newcommand{\mn}{{\mu\nu}}

 \newcommand{\diag}{{\rm diag}}

\newcommand{\ks}{{\rm ks}}

\newcommand{\cC}{{\cal C}}

\newcommand{\cE}{{\cal E}}

\newcommand{\cL}{{\cal L}}

\newcommand{\cLa}{{\cal L}}

\newcommand{\cO}{{\cal O}}

\newcommand{\cQ}{{\cal Q}}

\newcommand{\sth}{s_\theta}

\newcommand{\Gam}{\Gamma}

\newcommand{\Sn}{\Sn }

\newcommand{\tsh}{t}

\newcommand{\BEQ}{\begin{eqnarray}}   
\newcommand{\EEQ}{\end{eqnarray}}   
\newcommand{\BEA}{\begin{eqnarray}}   
\newcommand{\EEA}{\end{eqnarray}}   
\newcommand{\nn}{\nonumber }   
\renewcommand{\d}{{\rm d}}   

\renewcommand{\Sn}{N}
\newcommand{\Se}{S}

  \renewcommand{\thesection}{\arabic{section}}
   \renewcommand{\theequation}{\thesection.\arabic{equation}}
     \renewcommand{\thesection}{\arabic{section}}
 \renewcommand{\thesubsection}{\thesection\arabic{subsection}}
    \renewcommand{\theequation}{\thesection\arabic{equation}}
\renewcommand{\thefigure}{\thesection\arabic{figure}}

\begin{document}

 \title{The smeared-horizon observer of a black  hole}

\author{Theodorus Maria Nieuwenhuizen}

\affil*{\orgdiv{Institute for Theoretical Physics}, \orgname{University of Amsterdam}, 
\orgaddress{\street{Science Park 904}, \city{Amsterdam}, \postcode{1090 GL},  \country{ The Netherlands} \\ {t.m.nieuwenhuizen@uva.nl}}}




\abstract{
A class of observers is introduced that interpolate smoothly between the Schwarzschild observer, stable at spatial infinity,
and the Kerr-Schild observer, who falls into a black hole. 
For these observers the passing of the event and inner horizon takes a finite time, which diverges logarithmically 
when the interpolation parameter $\sigma$ goes to zero.
In the field theoretic approach to gravitation, the behavior at the horizons becomes regular,
making the mass of the metric well defined.   }

 
 
  
  \maketitle
 

 \tableofcontents
 
\vspace{3mm}
\hfill{Work dedicated to Igor V. Volovich} 

\section{Introduction}

\newcommand{\sh}{{\it sh}}
\newcommand{\ish}{{\it ish}}
\newcommand{\osh}{{\it osh}}

The most amazing property of black holes is their event horizon, which even
led Einstein initially to be skeptical about the Schwarzschild metric.
 An observer at spatial ``infinity'' will never live the day to observe something falling into a black hole. 
The related infinite redshift at the event horizon is the ultimate limit of the redshift 
known from sirenes on cars that  move away from us.

On the other hand, there are the Painlev\'e-Gullstrand and Kerr-Schild metrics, which do not exhibit a horizon
for the observers related to them. It is the aim of the present paper to introduce a class of intermediate observers,
the ``smeared horizon observers'', which interpolate between the Schwarzschild and Kerr-Schild metrics by a free parameter $\sigma$.
Some related properties are discussed.

In section 2 we introduce the metric of the smeared-horizon observers and consider properties like its eigenvectors and their behaviors
in the black hole interior, and analyze outgoing and ingoing spherical mass shells.
In section 3 we connect to recent exact solutions for the black hole interior and its role for the smeared-horizon observer.
In section 4 we connect to a recent class of exact solutions for smooth, cored black hole interiors.
In section 5  we show that the field theoretic approach to gravitation, connected to the Landau-Lifshitz pseudo tensor for
the gravitational field, becomes well defined at the would-be horizons and hence everywhere,
so that allows to properly define the mass of the metric.

\renewcommand{\thesection}{\arabic{section}}
\section{Generalized Schwarzschild and Reissner-Nordstr\"om metric}
\setcounter{equation}{0} \setcounter{figure}{0}
\renewcommand{\thesection}{\arabic{section}.}

For smooth functions $\Sn(r)$ and $\Se(r)$ we consider the generalization of the Schwarzschild metric
in spherical coordinates $r^\mu=(t,r,\theta,\phi)$, $\mu=0,1,2,3$,
\BEQ \label{gmnNS}
\d s^2=-\Sn ^2\bar \Se  \d t_\Schw^2+\frac{1}{\bar \Se  }\d r^2-r^2\d\Om^2=g_\mn \d r^\mu\d r^\nu,\qquad
 \bar \Se  =\Se  -1,
\EEQ
with $\d\Om=(\d\theta,\sin\theta\d\phi)$.
The Schwarzschild metric \cite{schwarzschild1916gravitationsfeld}  is described by $\Sn (r)=1$, $\Se  (r)=2GM/r$; the Reissner Nordstr\"om metric 
\cite{reissner1916eigengravitation,nordstrom1918energy} by $\Sn =1$,  $\Se  =G(2M/r-Q^2/r^2)$.
The latter has an event ($e$) horizon and an inner ($i$) horizon, where $\Se  (R_{e,i})=1$ ($\bar \Se  =0$). They are located at
\BEQ 
\hspace{-3mm}
R_e=G(M+\sqrt{M^2-\cQ^2}),\quad
R_i=G(M-\sqrt{M^2-\cQ^2}),\quad \cQ=m_PQ .
\EEQ
In our units $\hbar=c=1$ and  $\mu_0=4\pi$, the Planck mass is $m_P=1/\sqrt{G}$. 

It holds that $0<S<1$ in the outer space, so that $\bar S<0$ there. 
Inside a core bounded by the inner horizon $R_i$, one has $S<1$. In the Schwarzschild metric, the inner horizon coincides with
the origin, but this is a special case. In the Reissner Nordstr\"om metric,  $S\to-\infty$ for $r\to 0$. 

We have recently proposed a class of exact solutions where $S$ is regular with $S\sim r^2$ for $r\to0$. 
The latter property implies the presence of a finite core bounded by an inner horizon $R_i$.
As in the Reissner Nordstr\"om metric, the region between the inner and event horizons, termed  the {\it mantle}, is a standard vacuum,
described by  the Reissner Nordstr\"om metric with $S>1$.
In these models one has $N(r)=1$ for $r>R_i$ and $0<N\le 1$ for $r<R_i$.

\subsection{The Painlev\'e-Gullstrand observer}

\hfill{De \'e\'en heeft meer in zijn mars dan de ander\footnote{The one is more capable than the other}}

\hfill{Dutch expression}

\vspace{3mm}

One of the mysteries of black holes is that in its interior, from the event horizon to the inner horizon, the roles of $r$ and $t$ are interchanged.
The reversed role of $r$ and $t$ in its interior is counterintuitive, and so is the infinite redshift for signals from the horizon to a stationary observer at spatial infinity.

The first step to investigate the issue was  made by Painlev\'e in 1921 \cite{painleve1921mecanique}, 
and Gullstrand in 1922 \cite{gullstrand1922allgemeine}, actually intended to question the Schwarzschild 
solution. For completeness, we recall their approach.
In the Schwarzschild metric they introduce a new  time coordinate by setting
 $\d t_S=\d t_\pg -\sqrt{(2GM/r)} \,\d r/(1-2GM/r)$, so as to obtain
\BEQ
\d s^2=(1-\frac{2GM}{r})\d t_\pg^2 - 2 \sqrt{\frac{2GM}{r}}\d t_\pg\d r-\d r^2 -r^2\d\Om^2 .
\EEQ
This metric is regular except for $r\to0$, so that the observer does not notice an event horizon. 
In integral form one has  $t_S=t_\pg-2GM\{2y-\log[(y+1)/(y-1)]\}$, where $y=\sqrt{r/2GM}$.

 \subsection{The ingoing Kerr-Schild observer}

In this work we follow the approach of Kerr-Schild \cite{kerr1965some} to generate new metrics.
We start by allowing in the Kerr-Schild metric a dilated time $\d  t\to N(r) \d t_\ks$,
\BEQ \label{gmnKS}
\d s^2 &=&\Sn ^2(r)\d t_\ks^2-\d r^2-r^2\d\Om^2-\Se(r)  \d k^2(r) .
\EEQ
The one-form $\d k=k_\mu \d r^\mu$ involves a null vector $k_\mu$, viz. $g^\mn_\ks k_\mu k_\nu=0$.
The standard cases and their connection to $t_S$ are
\BEQ \label{S2PG}\label{S2iks}
&&\d k=k_\mu\d r^\mu=\Sn \d t_\ks+\d r, \quad   \d t_\Schw=\d t_\ks +\frac{\Se  }{\Sn \bar \Se  }\d r,\qquad  (\text{\it iks}), 
\\ & & \label{S2oks}
 \d k=k_\mu\d r^\mu=\Sn \d t_\ks-\d r,  \quad  \d t_\Schw=\d t_\ks -\frac{\Se  }{\Sn \bar \Se  }\d r, \qquad (\text{\it oks})
\EEQ
where $iks$ stands for an ingoing Kerr-Schild observer and $oks$ for an outgoing one.
Unlike (\ref{gmnNS}), this is regular also at $\Se  =1$, $\bar \Se  =0$. For the $iks$ case it reads
\BEQ
\d s^2 = -\Sn ^2\bar S \d t_\ks  ^2-2\Sn \Se  \d t_\ks \d r-(\Se  +1)\d r^2-r^2\d\Om^2 .
\EEQ
The inverse $iks$ metric is coded in
\BEQ
g^\mn\p_\mu\p_\nu= \frac{\Se  +1}{\bar \Se  \Sn ^2}\p_{t_\ks}^2 \mmin 2\frac{\Se  }{\Sn }\p_{t_\ks} \p_r  \pplus \bar \Se  \p_r^2 
-\frac{1}{r^2}\p_\Omega^2,\qquad \p_\Omega=(\p_\theta, \frac{1}{s_\th}\p_\phi) .
\EEQ
Indeed,  $k_\mu$ is a null vector, $k^\mu k_\mu=N^2g^{00}+2Ng^{01}+g^{11}=0$.

The situation for the $oks$ is obtained by setting $\d r \to -\d r$ and $\p_r\to-\p_r$.

\subsection{The ingoing smeared-horizon observer (\sho)}

The present work  proposes a new class of observers, to be called ``smeared-horizon observers'' ($sho$); the ingoing ones are 
for some $\sigma\ge0 $ defined by
\BEQ \label{dtSchwdtsh}
\d t_\ks = \d \tsh -  \frac{\Se  \bar \Se  }{\Sn (\sigma^2+\bar \Se  ^2)}\d r, \qquad
\d t_\Schw=\d \tsh+\frac{\sigma^2 }{\sigma^2+\bar \Se  ^2} \frac{\Se  }{\Sn \bar \Se  }\d r .
\EEQ
The latter form interpolates between Schwarzschild's stationary observer at infinity ($\sigma=0$) and the  
 ingoing PG observer  ($\sigma\to\infty$). This observer still falls in, but slower than the latter.

We shall take $\sigma$ constant, though it can actually be a function of $r$.

 The line element takes the form
\BEQ\label{gmneps}
\hspace{-5mm}
\d s^2=-\Sn ^2\bar \Se  \d \tsh ^2-2\frac{\sigma^2\Sn \Se  }{\sigma^2 \pplus \bar \Se  ^2}\d \tsh \d r 
\pplus (\bar \Se   \mmin \sigma^2)\frac{\sigma^2(\Se  \pplus 1)+\bar \Se  ^2}{(\sigma^2+\bar \Se  ^2)^2}\d r^2-r^2\d\Om^2,
\EEQ
with $g=\det(g_\mn)=-r^4\Sn ^2\sth^2$. The inverse metric is coded in
\BEQ \hspace{-3mm}
g^\mn\p_\mu\p_\nu=
\frac{\sigma^2-\bar \Se  }{\Sn ^2}\frac{\sigma^2(1+\Se  )+\bar \Se  ^2}{(\sigma^2+\bar \Se  ^2)^2}\p_{\tsh}^2
-\frac{2}{\Sn }\frac{\sigma^2\Se  }{\sigma^2+\bar \Se  ^2}\p_{\tsh} \p_r \pplus \bar \Se  \p_r^2
\mmin \frac{1}{r^2}  \p_\Om^2  . \EEQ
At any finite $\sigma$, $g_\mn$ and $g^\mn$ are regular, notably at the would-be horizon(s) where $\bar \Se  =0$.
Taking $\sigma$ small exposes Schwarzschild's event horizon at an arbitrary precision.

\subsection{Eigenbasis of the ingoing smeared-horizon metric}

The eigenvalues of $g_\mn^\ish$  in (\ref{gmneps}) are, next to $\lambda_2=-r^2$ and $\lambda_3=-r^2\sth^2$, 
     \BEQ
     \hspace{-3mm}
     \lambda_0= \Sn  e^{-2\cLa},\qquad      \lambda_1=-\Sn e^{2\cLa},\quad 
     \EEQ
     with $\cLa$ defined by
     \BEQ
     \sinh 2\cLa =\frac{ (\Se  +1) \sigma^4   -2\sigma^2\bar \Se   -\bar \Se  ^3 }{2\Sn  (\sigma^2+\bar \Se  ^2)^2} +\frac{\Sn \bar \Se  }{2}.
\EEQ
Outside the BH core, one has $\Sn =1$ and $\Se  =2GM/r-GQ^2/r^2$, so that $\cLa\to GM/r$ at large $r$.
In the limit $\sigma\to0$ one has $\sinh 2\cLa\to\half(\Sn ^2\bar \Se  ^2-1)/\Sn \bar \Se  $, which diverges at $\bar \Se  =0$
and changes sign there; at small $\sigma$, these divergences are rounded.
In that case, $\cLa$ changes sign at $\bar \Se  =\sigma^2+\cO(\sigma^4)$, which codes two
locations in the mantle, one beyond $R_i$ and the other below $R_e$.

The eigenvectors of the (0,1) sector of $g_\mn$ are mixed, 
\BEQ \hspace{-3mm} \label{eigvrs}
e_0 \! = \! \frac{(e^\cE \! ,-e^{-\cE},0,0)}{\sqrt{e^{2\cE}+e^{-2\cE}}},\hspace{6mm} e_1 \! = \! \frac{(e^{-\cE},\, e^\cE \! ,0,0)}{\sqrt{e^{2\cE}+e^{-2\cE}}} .
\EEQ 
while $e_2=(0,0,1,0)$ and $e_3=(0,0,0,1)$.
The parameter $\cE$ is defined by
\BEQ \label{shE=}
     \sinh2\cE &=& \frac{\sigma^2\pplus \bar \Se  ^2}{\sigma^2\Se  } \! \left(\sinh 2\cLa \mmin \Sn \bar \Se   \right)  \nn\\&=& \frac{\sigma^2\pplus \bar \Se  ^2}{\sigma^2\Se  } \!    \left ( \frac{ (\Se  +1) \sigma^4   -2\sigma^2\bar \Se   -\bar \Se  ^3 }{2\Sn  (\sigma^2+\bar \Se  ^2)^2} -\frac{\Sn \bar \Se  }{2} \right ) .
\EEQ
These identities show that $\cE=\cLa=\frac{1}{2}\log(\sqrt{2}+1)$ at the horizons where $N=1$, $\Se  =1$.
Eliminating $\Se  $ in favor of $\cE$,   the line element (\ref{gmneps}) takes the form
\BEQ
\d s^2= 
\Sn  e^{-2\cLa} \frac{(e^{\cE} \d \tsh-e^{-\cE}\d r)^2}{e^{2\cE}+e^{-2\cE}}
-\Sn  e^{2\cLa} \frac{(e^{\cE} \d  r+e^{-\cE}\d \tsh)^2}{e^{2\cE}+e^{-2\cE}}-r^2\d\Om^2.
\EEQ
Identifying this with the local Minkowski line element $\d\xi^0{}^2-\d\xi^1{}^2-\d\xi^2{}^2-\d\xi^3{}^2$, 
one  reads off  $\d\xi^2=r\d\theta$,  $\d\xi^3=r\sth\d\phi$ and the more interesting ones,
\BEQ \hspace{-3mm}
\d \xi^0=\sqrt{\Sn }\, e^{-\cLa}\, \frac{e^{\cE} \d t-e^{-\cE}\d r}{\sqrt{e^{2\cE}+e^{-2\cE}}} , \quad
\d \xi^1=\sqrt{\Sn }\, e^{ \cLa}  \, \frac{e^{\cE} \d r+e^{-\cE}\d t}{\sqrt{e^{2\cE}+e^{-2\cE}}} .
\EEQ
The signs are such that in the exterior, where $\cE\gg1$, $\d\xi^0\sim \d t$ and $\d\xi^1\sim\d r$. 
For  $\sigma\to\infty$  one describes the Kerr-Schild observer; with $\Sn \le1$ and $\Se\ge0$, $\cLa$ and $\cE$ remain positive,
making $t$ act as timelike coordinate also in the interior.

For small $\sigma$ there is a transition region which regularizes the singularity at the event horizon.
The regime $\bar \Se  <0$ applies to the outer space and also to the BH core;
in both cases, the prefactor $1/\sigma^2$ in $\sinh\cE$ makes $\cE \gg1$  and hence $\d\xi^0\sim\d t$ and $\d\xi^1\sim\d r$ as usual.
 In the mantle, the opposite case  $\bar \Se  >1$  involves $\sinh\cE<0$ and $\cE\ll-1$, so that there is the 
 switched connection between $t$ and $r$,  $\d\xi^0\sim - \d r$ and $\d\xi^1\sim\d t$, known from the interior of the Schwarzschild BH.
The width of the transition region is $\Delta \bar \Se  \sim\sigma^2$. 
For small $\sigma$ one can analyze the transition by coding $r$ in a parameter $\lambda$ such that
\BEQ
\bar \Se  =(1- \Sn \sinh \lambda)\sigma^2 +\cO(\sigma^4),\qquad
\EEQ
This yields  $\sinh 2\cE=\sinh 2\cLa+\cO(\sigma^2)=\sinh\lambda+\cO(\sigma^2)$. 
In the absence of surface layers, $\Sn = 1$ at the event and inner horizon, whence $r$ reads
\BEQ
r(\lambda)=G( M +  w) \big[1 +\frac{M + w}{2 w}  (\sinh\lambda-1) \sigma^2+\cO(\sigma^4)\big] ,\hspace{1mm} 
\EEQ
where $w=\pm \sqrt{M^2-\cQ^2}$, with the $+$ ($-$) sign at the event (inner) horizon.
Starting inside the mantle near the event horizon and going outwards, $\lambda$ increases from negative to positive values, 
with the event horizon $\bar \Se  =0$ located at $\lambda_c=\ln(\sqrt{2}+1)$.
Related behavior occurs around the inner horizon.


\subsection{Outgoing shells in the frame of the ingoing \sho}

An outgoing spherical shell for a massless field involves $\d\xi^0=\d\xi^1$, that is, 
\BEQ \label{drdtinobsin}
\frac{\d r}{\d t}=\frac{\sinh(\cE-\cLa)}{\cosh(\cE+\cLa)}.
\EEQ
This vanishes at the inner and event horizons where $\cE=\cLa$.
With $\Sn =1$ outside the core,  the large $r$ behaviors $\cLa\to GM/2r$ and $\sinh2\cE\to (1+\sigma^2)r/2GM\sigma^2$,
imply $\d r /\d t \to1-2GM\sigma^2/(1+\sigma^2)r$, an outgoing motion.

\myskip{ 
Near the event horizon one gets
\BEQ
\frac{\d r}{\d t}=\sigma^2(\sinh\lambda-1)\frac{\cosh\lambda }{1+e^{-\lambda}}
,\quad 2GM\frac{\d\lambda}{\d t}=\frac{\sinh\lambda-1 }{1+e^{-\lambda}}
\EEQ
The the real root of $\sinh\lambda_c=1$ is $\lambda_c=\log(\sqrt{2}+1)$.
In $t$-units of $2GM$,  this has the primitive
\BEQ
t=\log\vert \sinh\lambda-1\vert -\lambda+t_0+\log 2,\quad 
\EEQ
For $\lambda>\lambda_c$ 
\BEQ
\lambda=\log\frac{1+\sqrt{2-e^{t-t_0}}}{1-e^{t-t_0}}
\EEQ

For $t\to-\infty$, this reaches $\lambda_c$ where $\sinh\lambda_c-1=0$.
In this approximation, $\lambda\to\infty$ is reached for $t\uparrow t_0$.
The solution for $\lambda<\lambda_c$ is 
\BEQ
\lambda=\log\frac{1+\sqrt{2+e^{t-t_0}}}{1+e^{t-t_0}}.
\EEQ
For large positive $t$ this behaves as $\lambda\sim-\half  t$.
Increasing $\lambda$, and thus increasing $r$, corresponds to decreasing $t$, with $\lambda\uparrow \lambda_c$ 
for $t\to-\infty$.
The cases can be combined by setting 
\BEQ
\tau={\rm sg}(\lambda-\lambda_c)e^{t-t_0}
\EEQ
which is always increasing and passes through 0.  The results combine as
\BEQ
\lambda=\log\frac{1+\sqrt{2-\tau}}{1-\tau},\quad
\bar \Se  =-\sigma^2(\sinh\lambda-1)=-\sigma^2\frac{1+\sqrt{2-\tau} }{2(1-\tau)}\tau
\EEQ
which goes from $-\infty$ for $\tau\to-\infty$ to $+\infty$ for $\tau\to1$, passing $\lambda_c$ at $\tau=0$.
} 

Eq. (\ref{shE=}) shows that the crossing $\cE=\cLa$  occurs at the would-be horizons $R_{i,e}$ where $\Se  =1$.
Setting $t_e=-1/\Se  '(R_e) $, the  relation
\BEQ \hspace{-7mm}
\d t
=\frac{\cosh (\cE+\cLa)} {\sinh (\cE-\cLa) } \frac{\cE-\cL} {\cE'-\cLa'  }
\d \log\vert\cE-\cLa\vert\approx 2t_e\,\d\log \frac{\vert r-R_e\vert}{\sqrt{2}t_e} ,
\EEQ
leads for $r>R_e$ to $r-R_e\sim \exp(t/2t_e)$ and  for $r<R_e$ to $R_e-r\sim \exp(t/2t_e)$. The passing of the event horizon occurs for $t\to-\infty$.
Near the inner horizon we set $t_i=1/\Se  '(R_i) $ and obtain likewise
$r-R_i\sim \exp(-t/2t_i)$ and  for $r<R_i$ to $R_i-r\sim \exp(-t/2t_i)$; the passing occurs for $t\to+\infty$.
In the mantle, $t$ decreases when $r$ increases.


 Let us apply this to the Schwarzschild metric ($\Se  =2GM/r$ and $\Sn =1$), employ units $2GM \to 1$, so that $t_e\to1$, and define $\bar r=r-1$. One has
\BEQ
&&
e^t=(r-1)^2\,e^{f(r)}
 \nn\\ &&
f(r)=r+\frac{\sigma}{\sigma^2+1}\arctan\frac{\sigma^2r+\bar r }{\sigma}-\frac{\log(\sigma^2r^2+\bar r^2)}{2(\sigma^2+1)} .
\EEQ
It keeps the Schwarzschild singularity $t\sim \vert \log\bar r\vert $ for the time to go from a point $0<r-1\ll 1$ just outside of the event horizon to a location well away. 

Including a charge  $Q$, the adimensional units express $\Se  =2GM/r-GQ^2/r^2$ as $\Se  =1/r-q^2/4r^2$ 
with $q=m_PQ/M$. It yields
\newcommand{\rooti}{r_j}
\BEQ  \label{trootsumout}
&& 
t=
r+\Re\left(\frac{1-\half q^2}{\sqrt{1-q^2}} \log\frac{r-r_e}{r-r_i}+\log[(r-r_e)(r-r_i)]\right)+\Delta t,
  \nn\\ &&
  \Delta t=
\Re\sum_{j=1}^4
\frac{ q^4 - 8 q^2 \rooti  + 4(4 +q^2) \rooti ^2 - 16  \rooti ^3}{-q^2 + 2(2 +  q^2) \rooti  - 
 12 \rooti ^2 + 8(1+ \sigma^2) \rooti ^3} \frac{\log(r - \rooti )} {8}
\EEQ
where $r_{e,i}=\half(1\pm\sqrt{1-q^2})$ denote the event and inner horizons, respectively, and the $\rooti$ are the 4 complex roots of $\bar \Se  ^2+\sigma^2=0$.
Since they arise from $\bar \Se  =\pm i\sigma$, they take the explicit forms
\BEQ
r_{1,2}=\frac{1\pm \sqrt{1-q^2(1+ i\sigma)}}{2(1+ i\sigma)},\qquad
r_{3,4}=\frac{1\pm\sqrt{1-q^2(1- i\sigma)}}{2(1- i\sigma)} .
\EEQ
Since $r_{3,4}=r_{1,2}^\ast$, it follows that $\Delta t=2\Re(\Delta t_1+\Delta t_2)$ with
\BEQ \label{Deltat12}
\hspace{-6mm}
\Delta t_{1,2}= \frac{\mp [1 \pm \sqrt{1 - q^2 (1 + i \sigma)} \, ]^2}{8(1 + i \sigma)\sqrt{1 - q^2 (1 + i \sigma)}}
\log\left(r- \frac{1 \pm \sqrt{1 - q^2 (1 + i \sigma)}}{2 (1+ i \sigma)}\right) .
\EEQ
 
It is seen that $t$ keeps logarithmic divergences for signals emitted close to $r_{e,i}$. 
As above for $q=0$, the effect of a finite $\sigma$ is to double their prefactor. In other words, for $\sigma\to0$,
$\Delta t$ absorbs half of the logarithms of the first line in (\ref{trootsumout}).
But at finite $\sigma$, $\Delta t$ itself is regular for all real $r$.

In conclusion, in the description by the ingoing $sho$, it takes infinite time for the shell to emerge from the BH.
This stems with the popular statement ``nothing can escape from a black hole''.

\subsection{Ingoing shells described by the ingoing  {\it sho}}

An ingoing spherical shell for a massless field involves $\d\xi^0=-\d\xi^1$, that is, 
\BEQ
\frac{\d r}{\d t}=-\frac{\cosh(\cE-\cLa)}{\sinh(\cE+\cLa)} ,
\EEQ
which is finite for $\cE=\cLa$, that is, at the horizons.
Integrating this for the Schwarzschild metric yields
\BEQ &&
 t=-r+ \frac{\sigma}{\sigma^2+1} \arctan \frac{\sigma^2r+\bar r}{\sigma}  
-\frac{ \log(\sigma^2 r^2+\bar r^2)}{2 (\sigma^2+1)}  .
 \EEQ
 This remains finite as $t\sim \log1/\sigma$ for $r\to1$, so the ingoing massless shells are observed to go quickly through the event horizon.
For the charged case,  it involves $\Delta t$ of eq.  (\ref{trootsumout}),
\BEQ \label{trDeltat-ingoing}
t=-r +\Delta t .
\EEQ
Compared to the outgoing case (\ref{trootsumout}), next to $r\to-r$,  the explicit logarithms disappear, while the root sum is maintained. 

In conclusion, for the incoming \sho, incoming shells only need a finite time to pass the event horizon:
 for that process, no event horizon is noticed.

\section{The outgoing smeared-horizon observer }

So far we considered the ingoing PG observer.
The outgoing PG observer relates formally to the Schwarzschild observer by $r$-reversal in going from (\ref{S2PG}) to (\ref{S2oks}), so that
\BEQ \label{S2PGout}
\d t_\Schw=\d t_\ks -\frac{\Se  }{\Sn \bar \Se  }\d r
\EEQ
Correspondingly, there is a sign change in the shift term in (\ref{dtSchwdtsh}).
\BEQ \label{dtSchwdtshout}
\d t_\ks = \d \tsh +  \frac{\Se  \bar \Se  }{\Sn (\sigma^2+\bar \Se  ^2)}\d r, \qquad
\d t_\Schw=\d \tsh  - \frac{\sigma^2 }{\sigma^2+\bar \Se  ^2} \frac{\Se  }{\Sn \bar \Se  } \d r. 
\EEQ
The latter form interpolates between Schwarzschild's stationary observer at infinity ($\sigma=0$) and the  
outgoing KS observer  ($\sigma\to\infty$).

The reversed role of ingoing and outgoing motion corresponds to time-reversal, and turns the black hole into a so-called white hole.
In the exterior, $\d r/\d \tsh>0$ leads to time delay $\d t_S/\d \tsh>1$.
In the Schwarzschild interior, increasing time-like coordinate $r$ corresponds to increasing time.

With $ \d r\to-\d r$ and fixing the overall sign such that $e_0\to (1,0,0,0)$ and $e_1\to(0,1,0,0)$ for large $r$,
 the eigenvectors (\ref{eigvrs}) become
\BEQ \hspace{-3mm} \label{eigvrs3}
e_0 \! = \! \frac{(e^\cE \! ,e^{-\cE},0,0)}{\sqrt{e^{2\cE}+e^{-2\cE}}},\hspace{6mm} e_1 \! = \! \frac{(-e^{-\cE},\, e^\cE \! ,0,0)}{\sqrt{e^{2\cE}+e^{-2\cE}}}, \hspace{1mm}
\EEQ 
with $\cE$ again defined by (\ref{shE=}). This leads to
\BEQ \hspace{-3mm} \label{eigvrs2}
\d \xi^0=\sqrt{\Sn }\, e^{-\cLa}\, \frac{e^{\cE} \d t+e^{-\cE}\d r}{\sqrt{e^{2\cE}+e^{-2\cE}}} , \quad
\d \xi^1=\sqrt{\Sn }\, e^{ \cLa}  \, \frac{-e^{-\cE}\d t+e^{\cE} \d r}{\sqrt{e^{2\cE}+e^{-2\cE}}} .
\EEQ
In the exterior, one has $N=1$ and $\cE\gg 1$ for small $\sigma$,  so that $\d\xi_0=e^{-\cLa}\d t$ and $\d\xi_1=e^{ \cLa}  \d r$, as usual;
in the mantle  $\cE \ll -1$ so that  $\d\xi_0=e^{-\cLa}\d r$ and $\d\xi_1=-e^{ \cLa}  \d t$; and 
in the core  $\d\xi_0=\sqrt{\Sn }e^{-\cLa}\d t$ and $\d\xi_1=\sqrt{\Sn }e^{ \cLa}  \d r$. 

Outgoing shells are described by $\d \xi_1=\d\xi_0$, so that 
\BEQ \label{drdtoutout}
\frac{\d r}{\d t}=\frac{\sinh(\cL+\cE)}{\cosh(\cL-\cE)}
\EEQ
and ingoing ones by $\d \xi_1=-\d\xi_0$, so that 
\BEQ
\frac{\d r}{\d t}=\frac{\sinh(\cL-\cE)}{\cosh(\cL+\cE)}.
\EEQ
Apart from an overall sign, this coincides with (\ref{drdtinobsin}). Hence
ingoing shells, described by the outgoing \sho, take infinite time to pass the horizon.
Outgoing shells, on the other hand, only take a finite time, since (\ref{drdtoutout}) is regular.

\renewcommand{\thesection}{\arabic{section}}
\section{Exact solutions for the black  hole interior}
\renewcommand{\thesection}{\arabic{section}.}

A class of exact  solutions for the BH metric, which is regular everywhere in the core, was proposed recently.

\subsection{The stress energy tensor for the \sho}

For general $\Sn,\Se$ and $\sigma$, the Einstein tensor has the nontrivial elements
\BEQ  \label{EinstEq}
&& \hspace{-3mm}
G^0_{\,0}=\frac{\Se  +r\Se  '}{r^2},
\quad  G^1_{\,1}= \frac{ \Se  + r \Se  '}{r^2} +\frac{2 \Sn ' \bar \Se  }{r\Sn } , \quad
G^0_{\,1}=\frac{-2\sigma^2\Sn '\Se  }{ r\Sn ^2(\sigma^2+\bar \Se  ^2)} ,
 \nn\\ &&  \hspace{-3mm}
G^2_{\,2}=G^3_{\, 3}=\frac{2\Se  '+r\Se  ''}{2r} + \frac{\Sn '}{\Sn }\frac{2\bar \Se  +3r\Se  '}{2r} +\frac{\Sn ''}{\Sn } \bar \Se   , 
\EEQ
with $G^2_{\,2}=G^3_{\, 3}$ due to spherical symmetry. 
In the Schwarzschild case $\sigma=0$, $G^\mu_\ednu$ is diagonal; a value $\sigma>0$ does not modify these elements, but creates the $ G^0_{\,1}$ element provided $\Sn '\neq0$.
This represents a radial energy current for the smeared-horizon observer falling in onto the static energy distribution.


We express the stress energy tensor in terms of a local cosmological constant $\rho_\lambda$, an electrostatic energy density $\rho_E$
and thermal matter with velocity vector $U^\mu=\delta^\mu_0/\Sn \sqrt{-\bar \Se  }$  and stress energy tensor 
$T^\mu_{\vth\,\nu}=(\rho_\vth+p_\vth)U^\mu U_\nu-p_\vth\delta^\mu_\ednu$
involving an energy density $\rho_\vth$ and  isotropic pressure $p_\vth$.
The full stress energy tensor reads
 \BEQ \label{TMnrhomA}
\hspace{-6mm} 
T^\mu_\ednu
= (\rhom -p_\vth)\delta^\mu_\ednu+\rho_E \cC^\mu_\ednu
+(\rho_\vth+p_\vth) U^\mu U_\nu  , \quad 
\cC^\mu_\ednu =\diag(1,1,\mmin 1,\mmin 1) .
\EEQ
The Einstein equations $G^\mu_\ednu=8\pi G T^\mu_\ednu$ express the coefficients of (\ref{TMnrhomA})  in the functions $\Sn$ and $\Se$,   
\BEQ \label{EMT=}
\label{rhoL=}
&& \hspace{-6mm}
\bar\rho_\lambda=\frac{2\Se  \pplus 4r \Se  '\pplus r ^2\Se ''}{32\pi G r^2} \pplus 
\frac{\Sn ' }{\Sn }\frac{4 \bar \Se  \pplus 3 r \Se  '}{32\pi G r} \pplus \frac{\Sn ''}{\Sn }\frac{\bar \Se  }{16\pi G} , 
\\ &&  \hspace{-6mm}
\label{rhoA=}\label{EMTa=}
\bar\rho_E
=\frac{2 \Se  -r^2\Se  ''}{32 \pi  G r^2 }-\frac{\Sn ' }{\Sn }\frac{ 3\Se  ' }{32 \pi  G} -\frac{\Sn ''}{\Sn } \frac{\bar \Se  }{16 \pi  G }  ,
\\ &&  \hspace{-6mm}
\label{sigth=}
\bar\rho_\vth =-\frac{\Sn '\bar \Se  }{4\pi Gr\Sn }  , \qquad
\EEQ
where 
\BEQ
\bar\rho_\lambda\equiv \rho_\lambda +\frac{\rho_\vth-3p_\vth}{4}  ,\qquad
\bar\rho_E\equiv \rho_E +\frac{\rho_\vth+p_\vth}{4} ,\qquad
\bar\rho_\vth \equiv \rho_\vth+p_\vth . 
\EEQ
They combine as
\BEQ
\rho_\tot\equiv \rho_\lambda+\rho_E+\rho_\vth=\frac{\Se+r\Se'}{8\pi G r^2} .
\EEQ
While $U_0=1/U^0=\Sn \sqrt{-\bar \Se  }$ for any $\sigma$,  $U_1=-\sigma^2\Se  /(\sigma^2+\bar \Se  ^2)\sqrt{-\bar \Se  }$ 
connects the radial energy current $T^0_{\ed1}=G^0_{\ed1}/8\pi G$ from  (\ref{EinstEq}) to the thermal matter by $T^0_{\ed1}=\bar\rho_\vth U^0U_1$. 
Indeed, the $\rho_\lambda$ and $\rho_E$ terms of (\ref{EMTa=}) are proportional to the unit matrix in their (0,1) sectors, evidently for all $\sigma$. 
Hence the energy current relates to the real,  thermal particles; not to vacuum or electrostatic energy.


The electrostatic potential reads $A_\mu=\delta_\mu^0A_0(r)$. Conservation of the Maxwell tensor $F_\mn=\p_\mu A_\nu-\p_\nu A_\mu$ leads to
\BEQ
F^{\nu\mu}_{\td\,\, ; \nu}
=\left[A_0'\left(\frac{\Sn '}{\Sn ^3}-\frac{2}{r\Sn ^2}\right)-\frac{A_0''}{\Sn ^2}\right]\delta^\mu_0
=\mu_0 J^\mu .
\EEQ
With electric field $E(r)=-A_0'/\Sn $ and source $J^\mu=\delta^\mu_0\rho_q/\Sn $ this implies
\BEQ
E'+\frac{2}{r} E=\mu_0 \rho_q.
\EEQ
Taking $\mu_0=4\pi$, as usual in gravitation, this has the solution
\BEQ\label{EQrgoq}
E(r)=\frac{Q(r)}{r^2},\qquad Q(r)=4\pi\int_0^r\d u\, u^2\rho_q(u) .
\EEQ
So $Q(r)$ is  the enclosed charge. The related energy density is
\BEQ\label{rhoEEQ}
\rho_E=\frac{E^2}{2\mu_0}=\frac{Q^2(r)}{8\pi r^4}.
\EEQ
Notice that the connections (\ref{EQrgoq}) and (\ref{rhoEEQ}) are just as in special relativity.

Assuming that $\rho_\lambda$, $\rho_q$, $\rho_\vartheta$ and $p_\vth$ vanish in the mantle,
our task is to provide their physical meaning in the core $r\le R_i$, for suitable functions $\Sn$, $\Se$. 

\subsection{A class of exact solutions}
\label{exactsols}

We recently presented an exact solution for a charged  black hole core $r\le R_i$\cite{nieuwenhuizen2023exact}.
Here we recall it and then consider it for the smeared horizon observer.

The basic motivation is that in the stellar collapse, electrons are more easily ejected than protons, so that the black hole is 
positively charged. Since the Coulomb force is much stronger than the Newton force,
the fraction of surplus charge needs only be of order $m_N/m_P\sim 10^{-19}$.
The binding energy of the nucleons is released when their density is large enough.
The rest mass of the up and down quarks and the electrons makes up only $1\%$ of the energy;
when it is neglected, the problem allows an exact solution with $\Sn =1$, 
corresponding to a vanishing matter temperature and neglect of rest masses. 

Consider a core charge $Q_c$ be distributed as
 \BEQ
Q(r)=\qi F_q(\frac{r}{R_i}) =m_PR_i \,\qi  \, F_q(x) , \quad \qi  =\frac{\qi }{m_P R_i} ,
\quad x=\frac{r}{R_i} ,
\EEQ
with $F_q=1$ for $x\ge1$. This generates an electrostatic energy density
\BEQ\label{rhoAfq}
\rho_E=\frac{Q^2(r)}{8\pi r^4}=\frac{m_P^2}{8\pi R_i^2}  \frac{\qi ^2F_q^2(x)}{x^4} .
\EEQ
The solution for $\Se  (r)$ which goes from 0 at $r=0$ to 1 at $R_i$, is given by
\BEQ \label{S1exactgeneral} \hspace{-3mm}
\frac{\Se  (xR_i)}{x^2}=1+\frac{4}{3}\qi ^2\left[J(1)-J(x)\right],\quad
J(x)= \int_0^x \d y( \frac{1}{y^3}-\frac{1}{x^3})\frac{F_q^2(y)} {y^2}  ,
\EEQ 
with $0\le 	x\le1$. The integrals are well behaved when the charge density $\rho_q$ is finite at $r=0$, so that $F_q(y)\sim y^3$ for $y\to0$.
 
The solution rests on the insight that the zero point energy density of the quantum vacuum can act as a 
zero point battery or zero point storage, and locally absorb the energy density
\BEQ
\rho_\lambda =\frac{2\Se  +4r \Se  '+r ^2\Se  ''}{32\pi G r^2} ,\quad \qi ^2 =\frac{3}{1+4I_q},\quad 
I_q=\int_0^1\frac{\d x}{x^2}  F_q^2(x) .
\EEQ
For continuity with the vacuum, this must vanish at $R_i$, which  fixes $\qi$.
It holds that $s_i^- =R_i\Se  '(R_i^-) =\qi ^2 -1$, which should be non-negative at this first crossing of $\Se  =1$
starting from $\Se  =0$ at $r=0$.

Continuity $s_i^-=s_i^+=R_i\Se  '(R_i^+)=2( \Mc R_e/\qi ^2-1)$ sets 
\BEQ\label{Qc2Mc}
\frac{\cQc}{\Mc }=\frac{2\qi }{1+\qi ^2}
,\quad
R_i=\frac{2G\Mc }{1+\qi ^2}.
\EEQ
With $s_i$ between $0$ and $2$, $\cQc   /\Mc $ ranges from  $\half \sqrt{3}$ to $1$, that is to say, 
from quite charged to maximally charged. 

It has been put forward that surface charge layers may be present on the outer side of the inner and event 
horizons\cite{nieuwenhuizen2023exact}. 

The same idea of a nonuniform vacuum energy combined with electric fields has been applied to dark matter \cite{nieuwenhuizen2023solution}.

\subsection{The exact solutions for the smeared-horizon observer}

The connections (\ref{dtSchwdtsh}), (\ref{dtSchwdtshout})  lead to the transformation from $r^\mu_\Schw$ to $r^\mu_\sh$ given by
$\p r^\mu_\Schw/\p r^\nu_\sh=\delta^\mu_{\,\, \nu}+\alpha\delta ^\mu_0\delta^1_\nu$, 
with the inverse $\p r^\mu_\sh/\p r^\nu_\Schw=\delta^\mu_{\,\, \nu}-\alpha\delta ^\mu_0\delta^1_\nu$, where
$\alpha= \pm\,{\sigma^2 }\Se  /{(\sigma^2+\bar \Se  ^2}) {\Sn \bar \Se  }$.
The Einstein tensor transforms as 
$G^\mu_{\sh\,\nu}= (\p r^\mu_\sh/\p r^{\dot\mu}_\Schw) G^{\dot\mu}_{\,\,\dot\nu}(\p r^{\dot\nu}_\Schw/\p r^\nu_\sh)$,
which coincides with the diagonal $G^{\mu}_{\,\,\nu}$ tensor of the Schwarzschild case, 
and contains an extra term $G^0_{\sh\,1}=\alpha(G^0_{\,\,0}-G^1_{\,\,1})$, consistent with (\ref{EinstEq}).

The class of exact solutions of section \ref{exactsols} involves $\Sn (r)=1$ and hence $G^0_{\,\,0}=G^1_{\,\,1}$, so that the $G^0_{\,\,1}$  
term does not show up. But when $\rho_\vth$ and $p_\vth$ are nontrivial, so is $G^0_{\,\,1}$.
This relates already to the situation where thermal matter still involves a termperature $T=0$, but
 the rest masses of the up and down quarks and electrons are accounted for. They bring deviations at the percent level,
as shown in a numerical approach  \cite{nieuwenhuizen2023exact}.

\renewcommand{\thesection}{\arabic{section}}
\section{Einstein gravity as a field in flat space time}
\renewcommand{\thesection}{\arabic{section}.}

\myskip{
In this section I draw attention to the Landau-Lifshitz approach to gravitation, their pseudo tensor
for the gravitational part of the stress energy tensor, and the modest repair needed to make this a general 
field theoretic approach to gravitation, bringing gravitation at the same basis as (quantum) field theories in Minkowski space.
Finally, this approach is employed to evaluate the  BH mass as experienced by the various observers. }

The aim of the present section is to consider the mass of the regularized metrics of previous section.
When  the stress energy tensor $T^\mn$ is integrated over space, the conservation $T^\mn_{\td;\nu}=0$ 
does not lead to a conserved quantity. This led Landau and Lifshitz to derive their pseudo tensor $\tau^\mn$ representing 
the energy momentum tensor of the gravitational field itself\cite{landau1975classical}. 
However, it does not transform as a tensor. 
It has been pointed out that it becomes a proper object provided it is evaluated 
in Cartesian coordinates and transformed from there\cite{babak1999energy,nieuwenhuizen2007einstein}.
This approach is based on an underlying Minkowski space time, in which Noether's theorem assures a properly stress energy tensor.

\subsection{General approach}

Gravitation can be described as a field (a ``pudding'') in a standard Minkowski space, so that fits naturally with the matter fields in the standard model of particle physics.
We consider Cartesian and spherical coordinates,
\BEQ &&
\d \sigma^2=\eta_\mn \d x^\mu \d x^\nu=\gam_\mn \d r^\mu\d r^\nu,\quad x^\mu=(t,{\rm x},{\rm y},{\rm z}),\quad r^\mu=(t,r,\theta,\phi) ; \nn\\&&
\eta_\mn=\diag(1,-1,-1,-1),\quad \gam_\mn=(1,-1,-r^2,-r^2\sth^2),
\EEQ
and, for some  metric $g_\mn$,  the Riemann metric
\BEQ
\d s^2=g_\mn\d r^\mu\d r^\nu.
\EEQ
With  $g=\det(g_\mn)$ and $\gam=\det(\gam_\mn)$, the combinations
\BEQ 
k^\mn=\sqrt{\frac{g}{\gam}} g^\mn,\qquad
k_\mn=\sqrt{\frac{\gam}{g}} g_\mn,
\EEQ
act as tensor fields in flat space time.
This also applies to the mantle of the BH, which for the Schwarzschild metric is the full interior, 
even though the role of physical time is played there by the radial parameter $r$.

One can define the ``acceleration tensor''\cite{nieuwenhuizen2007einstein}
\BEQ
A^\mn =\half (k^\mn k^{\alpha\beta}-k^{\mu\alpha}k^{\nu\beta})_{:\alpha\beta} ,
\EEQ
where the column denotes covariant differentiation in flat space with its Christoffel coefficients $\gamma^\mu_{\ed\nu\rho}$ vanishing for Cartesian coordinates.
The Einstein equations take the form
\BEQ
A^\mn=8\pi G \Theta^\mn,\qquad \Theta^\mn
=\frac{g}{\gam}T^\mn+\theta^\mn=\frac{g}{\gam}(T^\mn+t^\mn).
\EEQ
$\Theta^\mn$ is conserved in Minkowski space, $\Theta^\mn_{\dd:\nu}=0$; this condition coincides 
with $T^\mn_{\dd;\nu}=0$, the conservation of $T^\mn$ in Riemann space.
By eliminating $T^\mn$ with use of the Einstein equations $G^\mn=8\pi GT^\mn$, one gets
\BEQ
8\pi G \theta^\mn=A^\mn-\frac{g}{\gam}G^\mn ,\qquad
8\pi G t^\mn=\frac{\gam}{g}A^\mn-G^\mn .
\EEQ
This expresses $t^\mn$ in terms of the metric alone. It can be verified that the second order derivatives drop out,
together with certain first order derivatives, so that there remains only a bilinear expression in first order derivatives,
\BEQ \label{tmn=}
&&
 t^\mn=\frac{1}{8\pi G}\big(
\half k^\mn_{\td:\lambda}k^{\lambda\rho}_{\td:\rho}
-\half k^{\mu\lambda}_{\td:\lambda}k^{\nu\rho}_{\td:\rho} 
+\half k_{\lambda\dot\lambda}k^{\rho\dot\rho}k^{\mu\lambda}_{\td:\rho}k^{\nu\dot\lambda}_{\td:\dot\rho} 
-\half k_{\lambda\dot\lambda}k^{\mu\dot\mu}k^{\lambda\rho}_{\td:\dot\mu}k^{\nu\dot\lambda}_{\td:\rho}
\nn\\&&
 -\half k_{\lambda\dot\lambda}k^{\nu\dot\nu}k^{\mu\lambda}_{\td:\rho}k^{\dot\lambda\rho}_{\td:\dot\nu}
+\frac{1}{4}k_{\lambda\dot\lambda}k_{\rho\dot\rho} k^{\mu\dot\mu}k^{\nu\dot\nu}
k^{\lambda\rho}_{\td:\dot\mu}k^{\dot\lambda\dot\rho}_{\td:\dot\nu}
-\frac{1}{8}k_{\lambda\dot\lambda}k_{\rho\dot\rho} k^{\mu\dot\mu}k^{\nu\dot\nu}
k^{\lambda\dot\lambda}_{\td:\dot\mu}k^{\rho\dot\rho}_{\td:\dot\nu}
\nn\\&&
+\frac{1}{4}k^\mn k_{\lambda\dot\lambda}  k^{\lambda\rho}_{\td:\alpha}k^{\dot\lambda\alpha}_{\td:\rho}
-\frac{1}{8}k^\mn k_{\lambda\dot\lambda}k_{\rho\dot\rho} k^{\alpha\dot\alpha}k^{\lambda\rho}_{\td:\alpha}k^{\dot\lambda\dot\rho}_{\td:\dot\alpha}
+ \frac{1}{16} k^\mn k_{\lambda\dot\lambda}k_{\rho\dot\rho} k^{\alpha\dot\alpha}
k^{\lambda\dot\lambda}_{\td:\alpha}k^{\rho\dot\rho}_{\td:\dot\alpha}\big).
\nn\\&& 
\EEQ

The fact that all second order derivatives
could be collected in $A_\mn$
arrives from absorbing the square-root factors in $k^\mn=\sqrt{g/\gam}\,g^\mn$.

In this field theoretic approach in Minkowski space, it is natural to identify $t^\mn$ as the stress energy tensor of the
gravitational field. In Cartesian coordinates it coincides with the Landau Lifshitz pseudo tensor. 
This clarifies its role: the Landau Lifshitz approach is correct in Cartesian coordinates; from them one can transform the 
results to any other coordinate systems. In the above approach, this is guaranteed by the covariant derivatives in flat space.
The material stress energy tensor in Minkowski space is $(g/\gam)T^\mn$.

 In the above cases, the determinants of the metrics are  $\gam=-r^4\sth^2$ and  $g=-\Sn ^2r^4\sth^2$, which
  implyies  $k^\mn=\Sn g^\mn$. The mass (energy) of the metric is
\BEQ \label{Eintegral}
E
=\int_{R^3}\d^3r\sqrt{-\gam}\Theta^{00} 
=\int_{R^3}\d^3r \, r^2\sin\theta\,\frac{A^{00}}{8\pi G},\quad \d^3r=\d r\d \theta\d \phi .
\EEQ 

For the Friedman metric in cosmology, it follows\cite{nieuwenhuizen2007einstein}  that $\Theta^{00}=0$, so that, 
loosely speaking, ``it costs no energy to create a universe''. This is the ultimate free lunch, more ultimate 
than the effect of the cosmological constant (dark energy, inflation) alone, 
for which the energy cost is known to be compensated by the gain of work.
Indeed, $\Theta^{00}=0$ holds also in the radiation and matter phases.

\subsection{The mass experienced by the smeared-horizon  observer}

General relativity allows various identifications of mass, in particular the far-field mass experienced by a
Newtonian observer. But a complete theory must deal with the near field and behaviour in the interior, and show that 
all is well there -- if it is. As we demonstrate now, this goal is reached for our class of smooth exact solutions observed  
by the smeared horizon observer.

For the metric (\ref{gmneps}), 
$\Theta^{00}$ is equal for the $\ish$ and $\osh$, reading 
\BEQ
\Theta^{00}&=& 
\frac{\sigma^6 (\Se  \pplus r \Se  ') \pplus \sigma^4 [\Se   (\Se  \mmin 3) \bar \Se   \mmin  r\Se  ' (3\Se  ^2\pplus 2\Se  \mmin 3)] }
{8\pi G r^2 (\sigma ^2+\bar \Se  ^2 )^3}  \nn\\ &&
-\frac{3 \sigma^2 \bar \Se  ^2 ( \Se   \bar \Se  \mmin r\Se  ')+   \bar \Se  ^4 ( \Se   \bar \Se  \mmin r\Se  ')  }
{8\pi G r^2 (\sigma ^2+\bar \Se  ^2 )^3}  , 
 \EEQ
 independent of $\Sn $. It is seen that $\sigma$ regularizes the $1/\bar S^2$ term, that is, the poles at the horizons.
 The other non-trivial elements of $\Theta^\mu_\ednu$, which is diagonal, are  independent of $\sigma$,
\BEQ
&& \Theta^1_{\ed1}=\Sn  \frac{ \Sn  (r \Se  ' + \Se  )+2 r \Sn ' \bar \Se   } {8 \pi G  r^2},\quad 
\\ 
&& \Theta^2_{\ed2}=\Theta^3_{\ed3}=\frac{\Sn ^2 (r \Se  '' + 2 \Se  ') + 2 r \Sn '^2 \bar \Se   + 
 2 \Sn  (r \Sn '' \bar \Se   + \Sn ' (1 + 2 \bar \Se   + 2 r \Se  '))}{6 \pi G r} . \nn
\EEQ

The Einstein equations yield for the material (non-gravitational) energy density 
\BEQ\label{TOO}
T^{00}=\frac{\sigma^2+\bar \Se  ^2+\sigma^2\Se   }{8 \pi G  r^2 \Sn ^2(\sigma^2 +    \bar \Se  ^2)^2} 
(\sigma^2-\bar \Se  ) (\Se  + r\Se  ').
\EEQ
The gravitational energy density $t^{00}=\Theta^{00}/\Sn ^2-T^{00} $ is proportional to $1/\Sn ^2$ and reads
\BEQ \label{tOO} 
 t^{00}&=&  -\frac{\bar \Se  ^5 \Se   \Se  ' }{2\pi G  r \Sn ^2 (\bar \Se  ^2 + \sigma^2)^3 }
-\frac{\Se   (\bar \Se  ^2 + \Se  ) + r (\bar \Se  ^2 - 7 \Se  ) \Se  ' }{8 \pi G  r^2 \Sn ^2(\bar \Se  ^2 + \sigma^2)^2 }   \bar \Se  ^3 
\nn \\ &&+ \frac{\Se   (\bar \Se  ^2 + \Se  ) + r (\bar \Se  ^2 - \Se  ) \Se  ' }{4 \pi G  r^2 \Sn ^2 (\bar \Se  ^2 + \sigma^2)} \bar \Se   
- \frac{1 + \Se  ^2 + r \Se   \Se  ' }{8 \pi G  r^2 \Sn ^2} .
\EEQ
The definition of $\Theta^\mn$ imposes that $\Theta^{00}$ is a total derivative. Indeed,
\BEQ\label{ThetaOO}
\Theta^{00}=\frac{1}{4\pi r^2}\frac{\d}{\d r}
 \frac{\sigma^4 - 2 \sigma^2 \bar \Se   - \bar \Se  ^3}{(\sigma^2 + \bar \Se  ^2)^2 }\frac{r\Se  }{2G}.
\EEQ
It leads to the integral
\BEQ\label{P0=}
P^{0}(r)=\int^r_0\d r\,4\pi r^2\,\Theta^{00}=
 \frac{\sigma^4 - 2 \sigma^2 \bar \Se   - \bar \Se  ^3}{(\sigma^2 + \bar \Se  ^2)^2 }\frac{r\Se  }{2G}.
\EEQ
The origin $r=0$ did not contribute, since $\Se  \sim r^2$ in our regularized approach.
Actually, the $r\to0$ value of (\ref{P0=}) vanishes even in the Schwarzschild case $\Se  =2GM/r$
and the RN case $\Se  =2GM/r-GQ^2/r^2$.

The important finding is that the quadratic divergencies of (\ref{ThetaOO}) in the $\sigma=0$ case, 
at the $\bar \Se  =0$ locations of the inner and event horizons, are regulated by any finite $\sigma$,
so that there is no longer a nasty  ``integration across the poles''.

With regularized $\Se  (r)\to0$ at $r\to0$, the mass is $E=P^0(\infty)$; since $\Se  =2GM/r-GQ^2/r^2$ for $r>R_e$, 
this  yields 
\BEQ 
E= M \text{    for  all    } \sigma.
\EEQ

For the Schwarzschild metric, the mass $M$ is in the field theoretic approach determined by the gravitational field alone, 
since $T^{00}=0$ implies that its energy density equals $t^{00}=\Theta^{00}$, while the singularity at $r\to0$ is
put under the rug. In terms of $\tilde r=r/2GM$ it reads when regularized by $\sigma$
\BEQ \label{tOOSchw}
\hspace{-4mm}
t^{00}&=&\frac{2 \tilde r -1 - 3 (2 + \sigma^2) \tilde r^2 +
  2 (2 + 3 \sigma^2 + 2 \sigma^4) \tilde r^3 -(1 + 3 \sigma^2 + 2 \sigma^4) \tilde r^4}
 {(8\pi Gr^2) \, [(\tilde r-1)^2  + \sigma^2 \tilde r^2]^3} \nn\\&& 
 =\frac{1}{4\pi r^2}\frac{\d}{\d r} \left\{
 \frac{ (\bar r-1)^2+ 2 \sigma \bar r^2}{[(\bar r-1)^2 + \sigma^2 \bar r^2]^2} (\bar r-1) M\right\}
\EEQ
That the integral is regular and finite, yielding $M$ as it should be, 
cannot hide that $\Theta^{00}$ and $t^{00}$ have an $1/r^2$ divergence at the origin,
pointing at singular behavior of the Schwarzschild metric.

Equations (\ref{ThetaOO}), (\ref{TOO}), (\ref{tOO}) and (\ref{tOOSchw}) show that the smeared horizon observer encounters for small 
$\sigma$ a smeared, non-sharp horizon, and only  a true one when the limit $\sigma\to0$ is taken first. 
The mass is thus well defined and takes the far field value $M$ for any finite value of $\sigma$.

\renewcommand{\thesection}{\arabic{section}}
\section{Outlook}
\renewcommand{\thesection}{\arabic{section}.}

The class of smeared horizon observers which are introduced in this papers allow for a complete and singularity-free description of black holes.
On the one hand, the exact solutions of ref. \cite{nieuwenhuizen2023exact} have been carried over without effort to these new observers;
on the other hand, their singularities at the inner and event horizons in the field theoretic description of gravitation, 
get smeared, so that the energy density is finite everywhere and the mass of the black hole well defined.

These smeared horizon observers keep some peculiarities:
for the ingoing observer, infalling shells cross the horizon in a finite time, but outgoing shells need infinite time.
For the outgoing observer, outgoing shells emerge in finite time, but infalling ones need an infinite time.

This puts forward the possibility to see matter falling into the core of a black hole in a finite time by an ingoing smeared horizon observer;
when this shell is next repelled by exerting some force on it, and made to go outwards, and, likewise, the observer is 
modified into an outgoing smeared horizon observer, he is capable to see the shell emerging  in a finite time.
A perhaps simpler setup is to investigate, for this type of observation, the geodesic motion of a point particle which enters the black hole mantle and next the core,
turns around the origin, and goes out again.

Another open question is the form of Hawking radiation for smeared-horizon observers.

\hspace{3mm} 

Data Availability Statement: No Data associated in the manuscript.


\begin{thebibliography}{11}
\ifx \bisbn   \undefined \def \bisbn  #1{ISBN #1}\fi
\ifx \binits  \undefined \def \binits#1{#1}\fi
\ifx \bauthor  \undefined \def \bauthor#1{#1}\fi
\ifx \batitle  \undefined \def \batitle#1{#1}\fi
\ifx \bjtitle  \undefined \def \bjtitle#1{#1}\fi
\ifx \bvolume  \undefined \def \bvolume#1{\textbf{#1}}\fi
\ifx \byear  \undefined \def \byear#1{#1}\fi
\ifx \bissue  \undefined \def \bissue#1{#1}\fi
\ifx \bfpage  \undefined \def \bfpage#1{#1}\fi
\ifx \blpage  \undefined \def \blpage #1{#1}\fi
\ifx \burl  \undefined \def \burl#1{\textsf{#1}}\fi
\ifx \doiurl  \undefined \def \doiurl#1{\url{https://doi.org/#1}}\fi
\ifx \betal  \undefined \def \betal{\textit{et al.}}\fi
\ifx \binstitute  \undefined \def \binstitute#1{#1}\fi
\ifx \binstitutionaled  \undefined \def \binstitutionaled#1{#1}\fi
\ifx \bctitle  \undefined \def \bctitle#1{#1}\fi
\ifx \beditor  \undefined \def \beditor#1{#1}\fi
\ifx \bpublisher  \undefined \def \bpublisher#1{#1}\fi
\ifx \bbtitle  \undefined \def \bbtitle#1{#1}\fi
\ifx \bedition  \undefined \def \bedition#1{#1}\fi
\ifx \bseriesno  \undefined \def \bseriesno#1{#1}\fi
\ifx \blocation  \undefined \def \blocation#1{#1}\fi
\ifx \bsertitle  \undefined \def \bsertitle#1{#1}\fi
\ifx \bsnm \undefined \def \bsnm#1{#1}\fi
\ifx \bsuffix \undefined \def \bsuffix#1{#1}\fi
\ifx \bparticle \undefined \def \bparticle#1{#1}\fi
\ifx \barticle \undefined \def \barticle#1{#1}\fi
\bibcommenthead
\ifx \bconfdate \undefined \def \bconfdate #1{#1}\fi
\ifx \botherref \undefined \def \botherref #1{#1}\fi
\ifx \url \undefined \def \url#1{\textsf{#1}}\fi
\ifx \bchapter \undefined \def \bchapter#1{#1}\fi
\ifx \bbook \undefined \def \bbook#1{#1}\fi
\ifx \bcomment \undefined \def \bcomment#1{#1}\fi
\ifx \oauthor \undefined \def \oauthor#1{#1}\fi
\ifx \citeauthoryear \undefined \def \citeauthoryear#1{#1}\fi
\ifx \endbibitem  \undefined \def \endbibitem {}\fi
\ifx \bconflocation  \undefined \def \bconflocation#1{#1}\fi
\ifx \arxivurl  \undefined \def \arxivurl#1{\textsf{#1}}\fi
\csname PreBibitemsHook\endcsname

\bibitem{schwarzschild1916gravitationsfeld}
\begin{botherref}
\oauthor{\bsnm{Schwarzschild}, \binits{K.}}:
{\"U}ber das {G}ravitationsfeld einer {K}ugel aus inkompressibler
  {F}l{\"u}ssigkeit nach der {E}insteinschen {T}heorie.
Sitzungsberichte der K\"oniglich Preu\ss ischen Akademie der Wissenschaften,
424--434
(1916)
\end{botherref}
\endbibitem

\bibitem{reissner1916eigengravitation}
\begin{barticle}
\bauthor{\bsnm{Reissner}, \binits{H.}}:
\batitle{{\"U}ber die {E}igengravitation des elektrischen {F}eldes nach der
  {E}insteinschen {}theorie}.
\bjtitle{Annalen der Physik}
\bvolume{355}(\bissue{9}),
\bfpage{106}--\blpage{120}
(\byear{1916})
\end{barticle}
\endbibitem

\bibitem{nordstrom1918energy}
\begin{barticle}
\bauthor{\bsnm{Nordstr{\"o}m}, \binits{G.}}:
\batitle{On the energy of the gravitation field in {E}instein's theory}.
\bjtitle{Bulletin van de {K}oninklijke {N}ederlandse {A}cademie van
  {W}etenschappen}
\bvolume{20},
\bfpage{1238}--\blpage{1245}
(\byear{1918})
\end{barticle}
\endbibitem

\bibitem{painleve1921mecanique}
\begin{barticle}
\bauthor{\bsnm{Painlev{\'e}}, \binits{P.}}:
\batitle{La m{\'e}canique classique et la th{\'e}orie de la relativit{\'e}}.
\bjtitle{Comptes {R}endus de l'{A}cad\'emie des {S}ciences de {P}aris}
\bvolume{173},
\bfpage{677}--\blpage{680}
(\byear{1921})
\end{barticle}
\endbibitem

\bibitem{gullstrand1922allgemeine}
\begin{bbook}
\bauthor{\bsnm{Gullstrand}, \binits{A.}}:
\bbtitle{Allgemeine {L}{\"o}sung des statischen {E}ink{\"o}rperproblems in der
  {E}insteinschen {G}ravitationstheorie}.
\bpublisher{Almqvist \& Wiksell}, \blocation{Stockholm}
(\byear{1922})
\end{bbook}
\endbibitem

\bibitem{kerr1965some}
\begin{bchapter}
\bauthor{\bsnm{Kerr}, \binits{R.}},
\bauthor{\bsnm{Schild}, \binits{A.}}:
\bctitle{Some algebraically degenerate solutions of {E}instein's gravitational
  field equations, applications of nonlinear partial differential equations in
  mathematical physics}.
In: \bbtitle{Proceedings of Symposia in Applied Mathematics},
vol. \bseriesno{17},
pp. \bfpage{199}--\blpage{209}
(\byear{1965})
\end{bchapter}
\endbibitem

\bibitem{nieuwenhuizen2023exact}
\begin{botherref}
\oauthor{\bsnm{Nieuwenhuizen}, \binits{T.M.}}:
Exact solutions for black holes with a smooth quantum core.
arXiv preprint arXiv:2302.14653
(2022)
\end{botherref}
\endbibitem

\bibitem{nieuwenhuizen2023solution}
\begin{botherref}
\oauthor{\bsnm{Nieuwenhuizen}, \binits{T.M.}}:
Solution of the dark matter riddle within standard model physics: From galaxies
  and clusters to cosmology.
arXiv preprint arXiv:2303.04637
(2023)
\end{botherref}
\endbibitem

\bibitem{landau1975classical}
\begin{botherref}
\oauthor{\bsnm{Landau}, \binits{L.D.}},
\oauthor{\bsnm{Lifshitz}, \binits{E.M.}}:
Classical field theory.
Course of Theoretical Physics
\textbf{2}
(1975)
\end{botherref}
\endbibitem

\bibitem{babak1999energy}
\begin{barticle}
\bauthor{\bsnm{Babak}, \binits{S.V.}},
\bauthor{\bsnm{Grishchuk}, \binits{L.P.}}:
\batitle{Energy-momentum tensor for the gravitational field}.
\bjtitle{Physical Review D}
\bvolume{61}(\bissue{2}),
\bfpage{024038}
(\byear{1999})
\end{barticle}
\endbibitem

\bibitem{nieuwenhuizen2007einstein}
\begin{barticle}
\bauthor{\bsnm{Nieuwenhuizen}, \binits{T.M.}}:
\batitle{{E}instein vs. {M}axwell: Is gravitation a curvature of space, a field
  in flat space, or both?}
\bjtitle{Europhysics Letters}
\bvolume{78}(\bissue{1}),
\bfpage{10010}
(\byear{2007})
\end{barticle}
\endbibitem

\end{thebibliography}
  
  

\end{document}